\documentclass[12pt]{iopart}

\usepackage{graphicx}
\begin{document}

\title{Constraining ultra light fermionic dark matter with Milky-Way observations}

\author{J. Barranco$^1$, Argelia Bernal$^{1,2}$, D. Delepine$^1$\footnote{Corresponding author}}
\address{$^1$ Departamento de F\'isica, DCI, Campus Le\'on, Universidad de Guanajuato, Le\'on, 37150, \\$^2$ Instituto de F\'isica y Matemáticas, Universidad Michoacana de San Nicol\'as de Hidalgo, Edificio C-3, Ciudad Universitaria, 58040 Morelia, Michoac\'an, México}

\ead{\mailto{jbarranc@fisica.ugto.mx},\mailto{bernal.a@ugto.mx},\mailto{delepine@fisica.ugto.mx} }
\vspace{10pt}
\begin{indented}
\item[]October 2024
\end{indented}

\begin{abstract}
The equation of state for a degenerate gas of fermions at zero temperature in the non-relativistic case is a polytrope, i.e. $p \sim\rho^{5/3}/m_F^{8/3}$. 
If dark matter is modeled by such a non-interacting fermion, this dependence in the mass of the fermion $m_F$ explains why if dark matter is very heavy the effective pressure of dark matter is negligible. 
Nevertheless, if the mass of the dark matter is very small, the effective pressure can be very large, and thus a system of self-gravitating fermions can be formed. 
In this work we model the dark matter halo of the Milky-Way by solving the Tolman-Oppenheimer-Volkoff equations,  with the equation of state for a partially degenerate ultralight non-interacting fermion.
We found that to fit the rotational velocity curve of the Milky-Way, the mass of the fermion should be in the range $31.5 ~\mbox{eV} < m_F < 35~$eV at $90\%$ C.L. 
Moreover, the central density is restricted to be in the range of  $1.2 < \rho_0<1.7$ GeV/cm$^3$ at $90\%$ C.L. 
The fermionic dark matter halo has a very different profile as compared with the standard Navarro-Frenk-White profile, thus, the possible indirect signals for annihilating dark matter may change by orders of magnitude. 
We found bounds for the annihilation cross section in this case by using the Saggitarius A* spectral energy distribution. 
\end{abstract}

\vspace{2pc}

\section{Introduction}
The determination of the properties of the particles that play the role of dark matter is perhaps the most active field in experimental and theoretical physics of our time \cite{Arcadi:2017kky,Bertone:2004pz,Jungman:1995df,ParticleDataGroup:2024cfk}. Despite this effort, up to now none of the properties of dark matter is known. Not even its mass or its spin. 
Considering only these two properties, the mass and the spin of the dark matter candidate, we can classify many models and candidates for dark matter in at least three main streams:
\begin{enumerate}
\item the heavy mass fermionic (spin one-half) candidate, i.e. the weakly interacting massive particle (WIMP) \cite{Lee:1977ua},\label{wimp}  
\item the ultra-light massive spin zero particle candidate, i.e. the axion-like particles  \cite{Weinberg:1977ma,Wilczek:1977pj,Peccei:1977hh,Matos:1998vk,Matos:2000ss,Hu:2000ke,Marsh:2015xka,Hui:2016ltb,Bernal:2010zz,UrenaLopez:2010ur,Bernal:2005hw,Robles:2018fur} and, \label{fuzzy}
\item the ultra-light fermionic  dark matter candidate, i.e. the sterile neutrino \cite{Dodelson:1993je,Shi:1998km,Dodelson:2005tp,Boyarsky:2018tvu,Dasgupta:2013zpn,Berezhiani:1995am,Randall:2015xza,Randall:2016bqw,Davoudiasl:2020uig}.
\label{esteril}
\end{enumerate}
The WIMP paradigm is strongly motivated by the elegant thermal freeze-out mechanism for dark matter production \cite{Lee:1977ua}. 
Among the many candidates for WIMP, the neutralino is the prototype together with other supersymmetric candidates \cite{Jungman:1995df}. 
Actually, the WIMP fits perfectly in the standard cosmological model because they are massive enough to be "cold relics". It means that they decoupled after they became non-relativistic. 
Astrophysical observations need in addition to cold dark matter (CDM) a cosmological constant $\Lambda$ to explain the present accelerated expansion of the universe. This is the so-called $\Lambda$-CDM model. 
Among their virtues, N-body simulations within a $\Lambda$-CDM universe fit most of the cosmological observables like Supernova Ia data, the power spectrum, weak lensing and more \cite{Aghanim:2018eyx}. 
Furthermore, it has some predictions: A universal density profile from hierarchical clustering for the dark matter halo of galaxies, commonly referred to as the Navarro-Frenk-White (NFW) dark matter profile \cite{Navarro:1996gj}. 
This profile has central densities that behave as $\rho\sim r^{-\beta}$, with $\beta \sim 1-1.5$. 
Nevertheless, observed galaxy rotation curves favor a constant density profile in contrast to the NFW profile. This is known as the {\it "Core-Cusp"} problem \cite{corecusp}. 
Moreover, the number of predicted satellite galaxies is bigger than the observed one. This problem is known as {\it the "missing satellite problem"} \cite{Klypin:1999uc}.
Another problem emerges from N-body simulations: the average density of dwarf galaxies is much higher than the observed densities of the local group, a problem known as the {\it Too big to fail} problem \cite{BoylanKolchin:2011de}.
More troubles arise as soon as current experimental efforts have not revealed any positive signal of existence of WIMPs. Neither by direct detection experiments \cite{Aprile:2018dbl,PandaX-4T:2021bab,LZ:2022lsv,XENON:2023cxc} or by
indirect detection \cite{Arcadi:2017kky,Chan:2019ptd,MAGIC:2016xys}.
Furthermore, DM collider production has not yet been observed \cite{Boveia:2018yeb}.   

To solve some of the galactic puzzles at small scales that permeate the WIMP paradigm, new proposals like fuzzy dark matter, sterile neutrinos or ultralight fermions have been introduced. 
Indeed, a plethora of ultra-light scalar candidates with properties similar to those of the axion have been postulated, called axion-like dark matter particles \cite{Weinberg:1977ma,Wilczek:1977pj,Peccei:1977hh,Marsh:2015xka}. 
Originally, the axion was proposed as a dynamical solution for the strong CP problem \cite{Weinberg:1977ma,Wilczek:1977pj,Peccei:1977hh}. It was soon realized that it could be a perfect dark matter candidate \cite{Ipser:1983mw} and for the last 40 years its mass and decay constant have been constrained. Surprisingly, it is still a viable candidate \cite{ParticleDataGroup:2024cfk}. 

In a similar approach as for axion or axion-like particles, it was proposed an ultra-light scalar particle with masses of the order of $m \sim 10^{-23}$ eV \cite{Matos:1998vk,Matos:2000ss,Hu:2000ke,Hui:2016ltb}. In contrast to axions or ALPs, the scalar particle is minimally coupled to gravity and has no other interactions with baryonic matter apart from the gravitational one. This proposal has been named fuzzy dark matter or scalar field dark matter (SFDM).

One virtue of this proposal is the emergence of a natural cut-off in the power spectrum \cite{Matos:2000ss}. Thus, no small halos exist in the SFDM model solving the "missing satellite" problem. Furthermore, in these models, the halos are described as self-gravitating systems composed of ultra-light scalar fields. In the particular case of complex scalar fields, these structures are known as Boson Stars (BS). For a scalar field mass of the order $m \sim 10^{-23}$ eV, the resulting BS have typical masses of $M \sim 10^{11} M_\odot$ and radii of several kiloparsecs, making them suitable as dark matter halo models. \cite{Bernal:2010zz,UrenaLopez:2010ur}. 

Furthermore, BS are regular at $r=0$ and thus, there is no "core-cusp" problem \cite{Bernal:2005hw}. In other words, ultra light bosonic particles might be free of some of the $\Lambda$-CDM problems. 
 However, it has been shown that, at small radii, within the SFDM model, more massive dark matter halos have denser cores than their CDM counterparts. Consequently, SFDM would require stronger baryonic feedback than CDM \cite{Robles:2018fur} to avoid the 'too big to fail' problem.

This motivates a revival on axion-like dark matter candidates, that is, particles that have some interaction with SM particles similar to the axion, although not necessary related with the strong CP problem, and with extreme low mass values, i.e $m \sim 10^{-23}-10^{-15}$ eV \cite{Marsh:2015xka,Hui:2016ltb}. 

Next, we arrive to the case of ultra-light fermionic  dark matter, i.e. particles with $1/2$ spin and masses between $\sim 1$ eV to $10$ KeV usually called sterile neutrinos \cite{Boyarsky:2018tvu,Shi:1998km,Dasgupta:2013zpn,Dodelson:1993je,Dodelson:2005tp}. Those DM candidates were advocated to improve predictions of small scale structure \cite{Dasgupta:2013zpn}. 
 The strongest constraints on sterile neutrino come from the Planck measurement on the number of relativistic species in the early Universe, $N_{eff}$, recently measured by the Planck collaboration \cite{Ade:2013zuv}. Those constraints can be evaded introducing a non standard relic density production mechanism \cite{Shi:1998km,Dasgupta:2013zpn}. Other constraints in these "warm" dark matter candidates come from Lyma-$\alpha$. Generally, all these constraints can be relaxed when dark matter is not in thermal contact with baryons. If this is the case,   when dark matter is decoupled from baryonic matter and DM is not reheated by interactions with other particles of the Standard Model, so, a light DM particle is cooler and non-relativistic earlier than a thermal relic, mitigating free streaming bounds \cite{Berezhiani:1995am,Randall:2015xza,Randall:2016bqw}. {Another possibility in order to relax Lyma-$\alpha$ constraints consists by adding more DM fermion species $N_f$ \cite{Randall:2016bqw,Davoudiasl:2020uig}. For the particular case of dark matter fermions with masses of the order of $\sim 1$ eV, this mechanism will require $N_f \simeq 10^4$ species. It is important to recall that the Lyma-$\alpha$ constraint assumes that the dark matter particles were once in kinetic equilibrium and have a specific thermal momentum distribution. The constraints can be ameliorated in scenarios where dark matter has a non-thermal momentum distribution. In the present work we are unable to provide a complete cosmological framework for a complete fermionic dark matter model that works in both scales, the cosmological and the galactic scales. }

Instead, following similar analyzes that have been done previously with ultralight fermions \cite{Randall:2016bqw,Destri:2012yn,Destri:2013pt,Domcke:2014kla}, here we will study the properties of dark matter halos when dark matter is an ultralight semi-degenerate fermion with no interaction with the standard model of particles except the pure gravitational one. { We will show that by using the data from our own galaxy, the model for dark matter that consists of a single ultralight fermion is severely constrained and that it will be difficult to explain different galaxies and consistent with a thermal cosmological model.} 
In this work, at least there are three main differences between previous works and the present one: i) we study for the first time the Milky-Way rotational curve data in order to constrain the mass of the fermions, ii) the fermion gas is treated as a semi-degenerate gas, allowing us to apply the formalism to large galaxies and iii) we obtain constraints on the annihilation cross section from the spectral energy distribution of the central black hole of the Milky-Way.  
An interesting property of ultra-light degenerate dark matter fermions is that even neglecting all other interactions, they have an intrinsic pressure that competes with gravitation. This pressure prevents fermions from collapsing gravitationally. In the standard $\Lambda$-CDM scenario, dark matter is pressureless. Thus, ultra-light fermions are expected to behave differently than standard WIMPs. A natural scenario where dark matter pressure may play a significant role is at galactic scales. In this scenario, ultra-light fermions can be treated as dark matter with pressure. For other works where dark matter with pressure has been studied at the galactic scales, see, for instance, \cite{Serra:2011jh, Barranco:2013wy, Boshkayev:2020vrg, Boshkayev:2021wns, Acena:2021wjx,Acena:2023qse}.

In a different approach, non-particle solutions to the "missing mass" problem in galaxies and galaxy clusters are possible as well. For instance, a radical alternative proposal is the Modified Newtonian dynamics (MOND) which was proposed in the 80's \cite{Milgrom:1983ca,Bekenstein:1984tv}. It proposes to replace dark matter entirely with a modified gravitational force law valid when accelerations reach a critical value  $a_0=1.2\times10^{-8}$cm/s$^2$. MOND successfully explains different phenomena at galactic scales, including the stellar dynamics in dwarf, elliptical and spiral galaxies (see, for instance, \cite{Sanders:2002pf} for a review) and can alleviate the Core-Cusp problem \cite{Re:2023qfu}. However, at the scale of galaxy clusters, it predicts radial temperature profiles that disagree badly with observations \cite{Aguirre:2001fj}. Other alternatives to solve the "missing mass" problem include cosmic backreaction. Backreaction has the potential to explain cosmic expansion and mimic dark matter without introducing any exotic dark matter component \cite{Vigneron:2019dpj}. Hypothetical black holes formed in the very early universe, named Primordial Black Holes (PBH), could potentially act as a significant component of dark matter \cite{Carr:2020xqk}. All these approaches are interesting on their own, although in a complex scenario, it is possible that a combination of all of them  can account for a complete description of dark matter phenomena.

In this work, we consider a simplified model in which the dark matter halo of the Milky Way is treated as a self-gravitating object composed of ultra-light fermions.
This is done by solving the Tolman-Oppenheimer-Volkoff equations with the equation of state for a partially degenerate ultralight non-interacting fermion.
It is found that in order to fit the rotational velocity curve of the Milky-Way, the mass of the fermion should be in the range $32.45 ~\mbox{eV} < m_F < 34.1~$eV at $68\%$ C.L. ($31.5 ~\mbox{eV} < m_F < 35~$eV at $90\%$ C.L. ). The effective pressure of ultra-light fermions leads to a non-cuspy dark matter profile with a regular central density. We found that the central density for the Milky-Way should be in the range of  $1.33~$GeV/cm$^3< \rho_0<1.58$ GeV/cm$^3$ at $68\%$ C.L. ($1.2~\rm{GeV/cm}^3 < \rho_0<1.7$ GeV/cm$^3$ at $90\%$ C.L.) to have consistency with the observed rotational curve of the galaxy. Because the central density is much smaller than a typical Navarro-Frenk-White dark matter profile, the possible indirect signals for annihilating dark matter may change by orders of magnitude.  We found bounds for the annihilation cross section in this case by using the Saggitarius A* spectral energy distribution. The spectral energy distribution of the central black hole is the emission of electromagnetic radiation produced by the accretion of matter into the suppermassive black hole at the center of the Milky-Way. This spectrum  covers different wavelengths and can be explained using physically motivated accretion models, such as relativistic magnetohydrodynamic  and advection-dominated accretion flow (ADAF) models \cite{Narayan:1997ku}. These models incorporate general relativity and plasma physics to simulate the emission from the accretion disk and jet. Various radiative processes are considered such as synchrotron, bremsstrahlung, and inverse Compton scattering of baryonic matter falling into the black hole. The models fit the observational data with good accuracy and without the need of incorporating dark matter emission. 

Our work and bounds differ from the so-called Ruffini-Argüelles-Rueda (RAR) model \cite{Ruffini:2014zfa,Arguelles:2018rjj,Krut:2023drx,Pelle:2024eyt} mainly because we indeed consider a supermassive black hole with mass of $4.1 \times 10^6M_\odot$ instead of a fermionic ball.  

The organization of the article is as follows: Typical self-gravitating objects made of ultra light fermions are presented in section \ref{section2}. Then, using galactic observations, namely the rotational velocity curve of the Milky-Way as reported by \cite{Sofue:2013kja,2019ApJ...871..120E}, the central density
$\rho_0$ and the mass of the fermion $m_F$ are constrained in section \ref{section3}. Furthermore, by using the spectral energy distribution (SED) of Saggitarius A* observed from the galactic center, new constraints on the annihilating cross section are obtained.  
Finally, some conclusions can be found in section \ref{section5}.

\section{Non-interacting fermionic dark matter halo} \label{section2}

A gas of fermions is degenerate at temperatures 
\begin{equation}
T<T_{Deg} =  \frac{\hbar^2}{2m_f K_B}\left(\frac{3\pi^2\rho}{ m_f}\right)^{2/3},
\end{equation}  
where $m_f$ is the mass of the fermion, $\rho$ the mass density of the fermion gas, and $K_B$ the Boltzmann constant. The mass density and pressure for a degenerate gas of fermions can be computed as \cite{Landau:1980mil,Narain:2006kx} \footnote{From this point we will use units where $c=\hbar=1$.}
\begin{eqnarray}
p&=&\frac{m_F^4}{24\pi^2}[(2z^3-3z)(1+z^2)^{1/2}+3\sinh^{-1}(z)]\,, \nonumber \\
\rho&=&\frac{m_F^4}{8\pi^2}[(2z^3+z)(1+z^2)^{1/2}-\sinh^{-1}(z)]\,. \label{eos1}
\end{eqnarray}
where $z=\frac{K_F}{m_F}$, $K_F$ the Fermi momentum.

There are two interesting limits, the non-relativistic case where $z\ll1$ (.i.e. the mass of the fermion is too big as
compared with the Fermi momentum) and the relativistic limit $z\gg1$.
In the non-relativistic case, by doing an expansion for $z\ll1$ in eqs. \ref{eos1} it is found that the pressure can be written as a function of the density. Indeed, it is found that
\begin{equation}
p=\frac{\left(3^2 \pi^4\right)^{1/3}}{5}\frac{\rho^{5/3}}{m_F^{8/3}}\,.\label{NR}
\end{equation}

It is interesting to note that the resulting equation of state for a gas of degenerate fermions is a polytrope, and it has only one free parameter: the mass of the fermion. 

In order to find the self-gravitating structure that a gas of free fermions may form, it is needed to solve Einstein's equations. As we have mentioned, 
in the non-relativistic limit, this gas of fermions satisfies an equation of state in a very similar way as a perfect fluid. 
Thus, the energy tensor that acts as a source for the Einstein equations can be written as:
\begin{equation}
T_{\mu\nu}=pg_{\mu\nu}+(p+\rho)U_\mu U_\nu\,.
\end{equation}
Here, $U_\mu$ is the four velocity of the fluid, $p$ and $\rho$ the pressure and density of the fluid.

For simplicity, we will restrict to the case of spherical symmetry and static space-time, that is, the Schwarzschild metric. 
 Spherical symmetry for the Milky-Way dark matter halo is well justified by the observation of carbon giant stars in the Galactic halo. Many of them are  clustered on a great circle on the sky that intersects the center of Sagittarius dwarf galaxy. That the stream is observed as a great circle indicates that the Milky-Way does not exert a significant torque on them, so the Galactic potential must be nearly spherical \cite{Ibata:2000pu}.

In this case, the
only unknown in the metric is the mass of the self-gravitating object. 
Imposing the condition of hydrostatic equilibrium one arrives to the following equations
\begin{eqnarray}
\frac{dp}{dr}&=&-\frac{GM\rho}{r^2}\left(1+\frac{p}{\rho}\right)\left(1+\frac{4\pi r^3p}{M}\right)\left(1-\frac{2GM}{r}\right)^{-1}\nonumber\,,\\
\frac{dM}{dr}&=&4\pi r^2\rho\, \label{mprima}.
\end{eqnarray} 
Those are the Tolman-Oppenheimer-Volkoff equations (TOV system). 

If the self-gravitating configurations  have low compactness, that is $GM/r \ll 1$, thus, the TOV system  is rewritten in the limit of low compactness as:
\begin{eqnarray}
\frac{dp}{dr}&=&-\frac{GM\rho}{r^2}\label{pressure}\,,\\
\frac{dM}{dr}&=&4\pi r^2\rho\label{mass}\,.
\end{eqnarray}
 These is  known as the the Newtonian limit of the TOV system.
 
The system can be solved once the equation of state (EOS) of the fluid is known. In the case of degenerate fermions, the corresponding EOS will be given by eq. \ref{NR}. Nevertheless, it could happen that at some point the temperature of the gas of fermions can be bigger than $T_{Deg}$. 
The temperature at any point of the gas can be determined using the virial theorem: 
\begin{equation}
T(r)=\frac{2}{3}\frac{G M(r) m_F}{K_B r}\,,
\end{equation}
where $K_B$ is the Boltzmann constant, $M$ is the enclosed mass of the compact object, and $r$ the radial distance where the temperature is computed.  
If $T > T_{Deg}$  eq. \ref{NR} is not valid anymore since it is valid only for a fully degenerate gas of fermions. Instead, the classical pressure will be used and it is given by:
\begin{equation}
P(\rho)=\frac{G M \rho}{2 r}\,.\label{classical}
\end{equation}
With those considerations, depending on the temperature of the fermion, the equation of state will be either the eq. \ref{NR} if the gas of fermions has a temperature below $T_{Deg}$ or equation of state eq. \ref{classical} if $T > T_{Deg}$.  

Consequently, the Newtonian TOV system we need to solve will depend on the corresponding temperature.
If $T < T_{Deg}$, the TOV system reduces to
 
 \begin{equation}
  \frac{d\rho}{dr} =-\left(\frac{3}{\pi^4}\right)^{1/3}\frac{GM}{r^2}m_F^{8/3}\rho^{1/3} \,, \quad \frac{dM}{dr}=4\pi r^2\rho \label{rhoprima}
 \end{equation} 
 and  if  $T  > T_{Deg}$ 
\begin{equation}
  \frac{d\rho}{dr} =-\frac{\rho}{r}-4\pi\frac{\rho^2 r^2}{M} \,, \quad \frac{dM}{dr}=4\pi r^2\rho.\label{rhoprima2}
 \end{equation} 

\begin{figure}[t]  %
\centering
\includegraphics[scale=0.4]{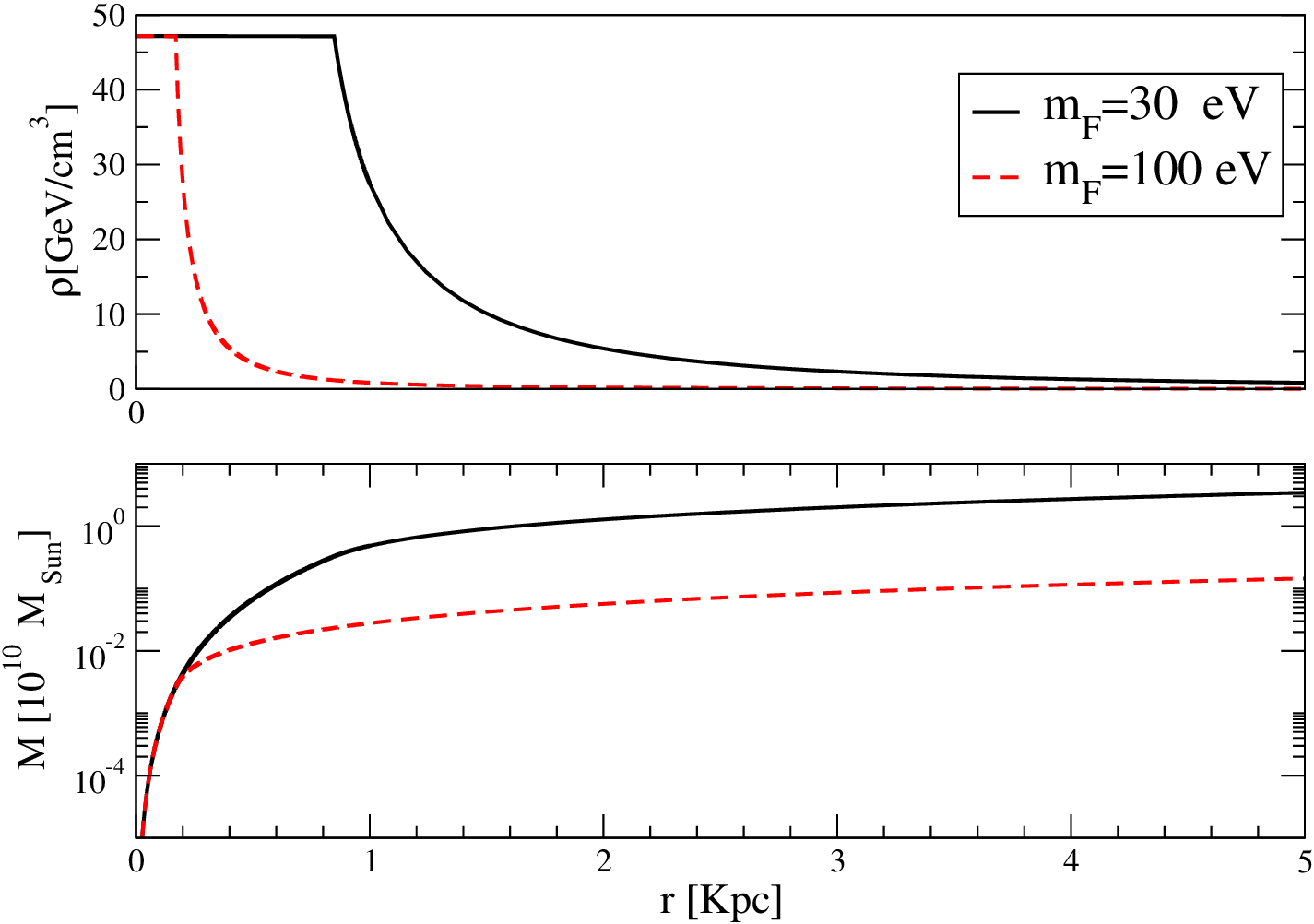}
\caption{ Self-gravitating structures made of non-interacting, non-relativistic semi-degenerate fermions at zero temperature. In the upper panel it is shown typical density and mass profiles for a central density of $\rho_0=47 ~\mbox{GeV/cm}^3$ for two different masses of the fermion, namely $m_F=30$ eV (solid) and $m_F=100$ eV (dashed). The flat central region of the density profile correspond to the degenerate gas. Lower panel shows the mass function.}\label{Fig1}
\end{figure}
\begin{figure}[t]  %
\centering
\includegraphics[scale=0.3]{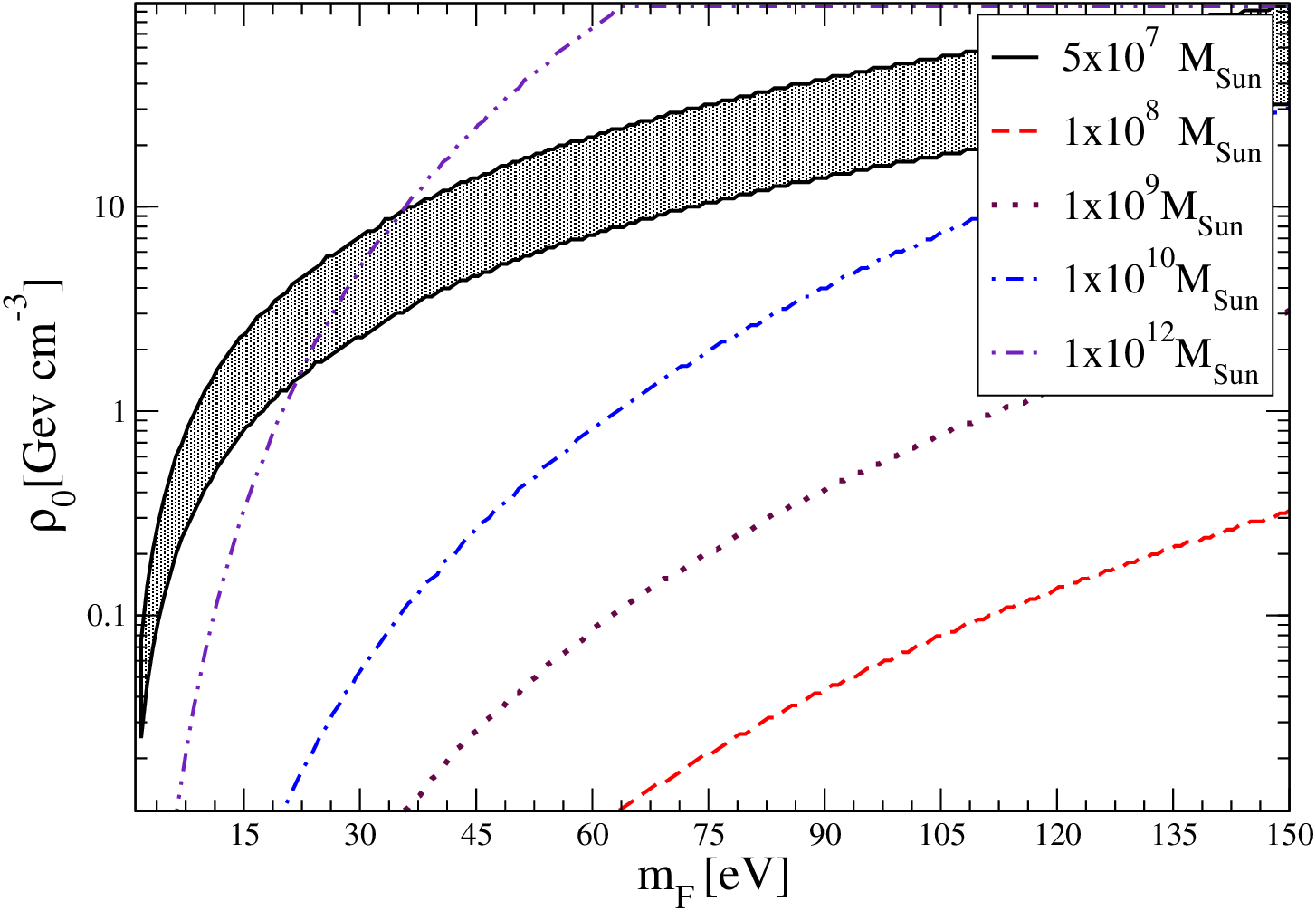}
\includegraphics[scale=0.3]{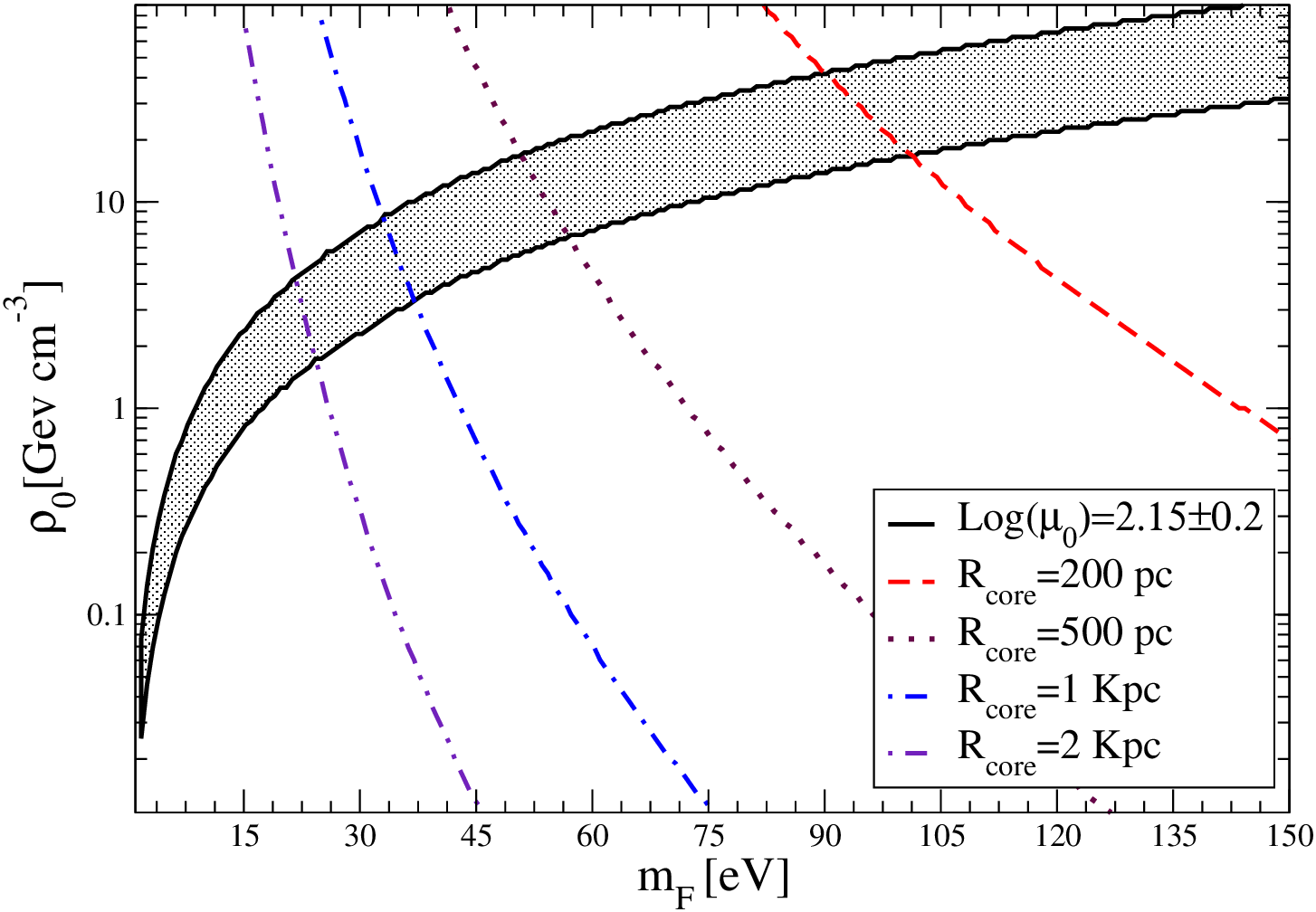}
\caption{Iso-curves of the total mass  $M(\rho_0,m_F)$ (left) and the radius of the core $R_{Core}(\rho_0,m_F)$ (right) for the full set of self-gravitating configurations obtained by varying the free parameters $m_F$ and $\rho_0$. The gray band corresponds to the allowed region for $\rho_0$ and $m_F$ needed to have configurations that fullfill the condition of constant dark matter surface density  eq. \ref{salucci}.}\label{Fig2}
\end{figure}

The equations \ref{rhoprima} can be solved numerically with boundary conditions $M(r=0)=0$ and $\rho(r=0)=\rho_0$. Thus $\rho_0$ is a free parameter. The integration of equations \ref{rhoprima} is performed until the point $T=T_{Deg}$. The radius $R_{core}$ where $T=T_{Deg}$ fixes the core of the configuration. At this point, the contained mass will be denoted as $M_*=M(R_{core})$ and will
correspond to a fixed mass density $\rho_*=\rho(R_{core})$. 
Now we can solve for $T>T_{Deg}$. The solution of equation \ref{rhoprima2} with boundary conditions 
$M(r=R_{core})=M_*$  and $\rho(r=R_{core})=\rho_*$ can be found exactly and they are given by:
\begin{equation}
\rho(r)=\frac{\rho_* M_* R_{core}}{r \sqrt{4 \pi\rho_* M_* R_{core} (r^2-R_{core}^2)+M_*^2}}\label{rhoafter}
\end{equation}
and
\begin{equation}
M(r)=\sqrt{4\pi\rho_* M_*R_{core}(r^2-R_{core}^2)+M_*^2}\,.\label{massafter}
\end{equation}
 Note that the mass density $\rho(r)$ decays as $r^{-2}$ and the mass goes as $M(r) \sim r$. The mass of the configuration is always a growing function of $r$ and the compact structure does not have a finite radius. Nevertheless, since $\rho(r)$ is a decreasing function for all $r$, we can define the radius of the self-gravitating object  as the point where  $\rho(r=R)<\rho_c$, with $\rho_c$ the critical density of the universe.  

The value of $R_{core}$ where $T=T_{Deg}$ is determined by $\rho_0$. Then, the self-gravitating system  have 
only two free parameters: 
\begin{itemize} 
\item The mass of the dark matter fermion $m_F$ and,
\item  the central density of the configuration $\rho_0$.
\end{itemize}

Typical configurations will have a density profile and a metric coefficient $M(r)$ as shown in Fig. \ref{Fig1}. For definitiveness,   two different values of the mass of the fermion are shown, namely, $m_F=30$ eV and $m_F=100$ eV for central density of $\rho_0=47~ \mbox{GeV/cm}^3$. Note that the masses of the self gravitating objects are of the order of  $M \sim 10^{11}~M_\odot$  and radii of $R \sim 10~$kpc for those values of the fermion mass. Thus it is clear that this self-gravitating structures made of semi-degenerate fermions might play the role for cored dark matter halos.
In what follows, we will say that the dark matter particle candidate is such non interactive fermion with ultralight values of the mass ($\sim$ few electron-Volts). 

In order to see the properties of the dark matter halos made of a gas of semi-degenerate ultralight fermions, we have constructed the full set of solutions of equations \ref{mass} -\ref{rhoprima2} obtained by varying the two free parameters of the dark matter halo: the central density $\rho_0$ and the mass of the fermion $m_F$.  

In Fig. \ref{Fig2} are shown iso-curves  of the total mass of the self-gravitating configuration $M(m_F,\rho_0)$. 

We can observe that for low central densities, the resulting configurations have properties that might explain dwarf galaxies data: core radius of the order of hundred parsecs and masses of the order of $\sim 10^{8}M_\odot$ for values of the fermion mass of $m_F \sim 100$ eV.  Similar results have been obtained in \cite{Randall:2016bqw,Domcke:2014kla} where the dispersion velocities from dwarf galaxies have been fitted resulting in upper limits on the mass of the dark matter fermion. 

On the other hand, for smaller masses of the fermions, the configurations might be suitable to explain larger galaxies, like elliptical or spiral  galaxies because the resulting core radius are the order of few kilo-parsecs, rotational velocities of hundreds of $km/sec$ and total masses of $10^{11}M_\odot$ for masses of the fermion of tens of eV. This is a consequence of the fact that the pressure of the fermion gas is inverse proportionality to the mass $m_F$.

The so called $\Lambda$-CDM model describes the large scale structure of the universe. Nevertheless, as previously mentioned, the $\Lambda$-CDM predictions at low scales are in debate \cite{Weinberg:2013aya}. In particular, the "core-cusp problem" \cite{corecusp} and the "too big to fail" \cite{BoylanKolchin:2011de} problems can be solved by invoking new interactions in the dark sector. Indeed, if dark matter has strong self-interactions, elastic scattering in the dense central region of halos will redistribute the energy and angular momentum among particles creating a core \cite{Spergel:1999mh}. Other solutions rely on supernova feedback and low star-formation efficiency. The first one might flatten the central cusp in big galaxies and a combination of both could explain why most of the Milky-Way's dark matter sub-halos do not host visible galaxies \cite{Governato:2012fa}. 

It is interesting that in our case, even without the addition of any other interaction in the dark sector, the resulting self-gravitating configurations might alleviate some of the $\Lambda$-CDM problems: There is no core-cusp problem since the dark matter halos have naturally a core $R_{core}$ produced by the degenerate gas. Furthermore, here there is not a "Too big to fail" problem since the central density $\rho_0$ is smaller than the central densities needed by NFW dark matter halos. 

There are some interesting scaling relations found empirically for the dark matter halos. In particular, in \cite{Donato:2009ab,Gentile:2009bw} it was found that the central surface density of galaxy dark matter halos defined as $\mu_{0D}=R_{core} \rho_0$, where $R_{core}$ and $\rho_0$ are the halo core radius and central density, respectively, is nearly constant. It was found \cite{Donato:2009ab,Gentile:2009bw}
\begin{equation}
\log\left(\frac{\mu_{0D}}{\mbox{M}_\odot \mbox{pc}^{-2}}\right) = 2.15 \pm 0.2 \label{salucci}
\end{equation}
and independent of galaxy luminosity. 

From all possible configurations for dark matter halos made of semi-degenerate fermions, there is a subset that can full fill that relation between core radius and central density. 

Those configurations defined a region in the $(\rho_0,m_F)$ parameter space that satisfies the condition eq. \ref{salucci}. This region is shown in the Fig. \ref{Fig2} as a gray band.   
Notice that this condition is fulfilled by a small fraction of the total possible self-gravitating configurations. We will show that if the Milky-Way dark matter halo is made of ultralight fermions, 
the resulting configuration do not satisfy eq. \ref{salucci}.  
\section{Milky-Way constraints}\label{section3}
\begin{table}
\begin{center}
\begin{tabular}{ccc}
\hline
Region (R)              & Central density  $\rho_R^c$                                         & Scale radius $a_R$\\
                               & $[10^{10} M_\odot \mbox{kpc}^{-3}]$                             & $[\mbox{kpc}]$ \\
\hline                                
Inner Bulge (IB)      & $\rho^c_{IB}=3.6\times 10^{3}$ & $a_{IB}=3.8\times10^{-3}$ \\
Main Bulge (MB)     & $\rho^c_{MB}=19.0$ & $a_{MB}=1.2\times10^{-1}$ \\  
Disk (D)     & $\rho^c_{D}=1.50$ & $a_{D}=1.2$ \\  
\hline
\end{tabular}
\end{center}
\caption{Values used to fit the inner region of the Milky-Way's rotational velocity curve}\label{table1}
\end{table}
From our results of the previous section, we observe that self-gravitating configurations for non interacting fermions cover a wide range of masses and core radii. Thus, they can be used to model different types of galaxies. From the smallest dwarf galaxies with $M \sim 10^7 M_\odot$ and core radius of some hundreds of parsecs up to spiral galaxies with $M \sim 10^{11} M_\odot$ and core radius of few kilo-parsecs. 

In this section we will use data form our own galaxy in order to see if this very simple model might fit the observed data. In contrast with other galaxies, our Milky-Way is perhaps one of the best known galaxies. We constrain the mass of the fermion $m_F$ and the central density $\rho_0$ with the rotational velocity of the stars in our galaxy. Then, we constrain the annihilation cross section for this ultra light candidate using the spectral information of Saggitarius A*. 

\subsection{Fermion mass constraints from rotational velocity curve}

\begin{figure}[t]  %
\centering
\includegraphics[scale=0.4]{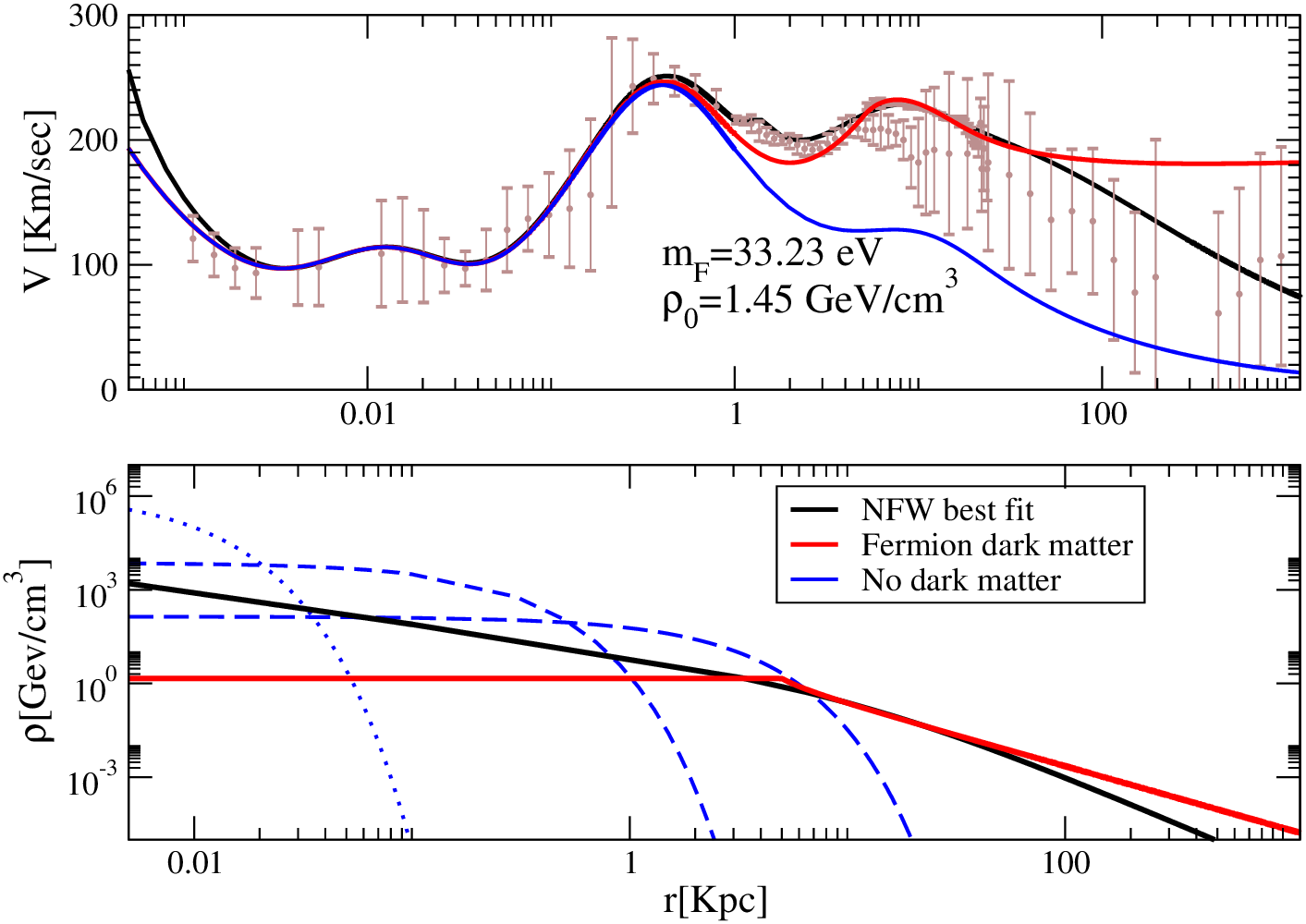}
\caption{Milky-Way rotational velocities and theoretical rotational curve  obtained for the best fit $\rho_0=1.45~\mbox{GeV cm}^{-3},m_F=33.23~\mbox{eV}$ for a dark matter halo model made of semi-degenerate non-interacting fermions )}\label{Fig3}
\end{figure}

Recent observation in the CO and CS line emissions from the central region of the galaxy have been used to derive the central rotational curve of the Milky-Way. 
This rotational curve covers a wide range of radius, from $\sim 1$pc up to several hundred kilo-parsecs \cite{Sofue:2013kja,2019ApJ...871..120E}. { Improved values in the region between 5 Kpc up to 25 Kpc are reported in \cite{2019ApJ...871..120E} performed by the GAIA collaboration.}

The inner rotational curve can be fitted if Milky-Way is divided in five different components:
\begin{enumerate}
\item The central black hole
\item A inner bulge or core of the galaxy
\item The main bulge 
\item The disk of the galaxy
\item A dark matter halo
\end{enumerate}
Each component can be modeled if the mass density is described by an exponential sphere model \cite{Sofue:2013kja,2019ApJ...871..120E}. In this case, the mass density profile $\rho(r)$ is given by:
\begin{equation}
\rho_R(r)=\rho_R^c\exp(-r/a_R)\,, \label{massdensity}
\end{equation} 
$a_R$ being a scale radius, $\rho_R^c$ the central density for each region, and $R$ a label to identify each of the regions that fits the Milky-Way. 

{ For the disk, in the approximation of a thin disk the circular rotational velocity cam be computed as \cite{Freeman:1970mx}:
\begin{equation}
v_{D}^2(r)=\frac{G M_{D}}{2 R_{D}}\left(\frac{r}{R_{D}}\right)^2\left[
I_0\left(\frac{r}{2R_{D}}\right)K_0\left(\frac{r}{2R_{D}}\right)-I_1\left(\frac{r}{2R_{D}}\right)K_1\left(\frac{r}{2R_{D}}\right)\right]\,,
\end{equation}
where $I_0,K_0,I_1$ and $K_1$ are the modified Bessel functions, $M_D$ is the total mass of the disk 
and for the Milky Way $M_D=4.4\times 10^{10}M_\odot$. Furthermore, $R_D$ is the scale length of the disk and for the Milky Way it is found to be $R_D=3$ Kpc.
}

The dark matter halo will be modeled by the self-gravitating structure of semi-degenerate fermions.

The values used to fit the rotational curve are shown in Table \ref{table1}. The existence of a supermassive black hole at the center of our galaxy is strongly supported by the motion of the S-star galaxies. Its mass has been constrained to be $M_{BH}=(4.1\pm0.6)\times10^6M_\odot$ \cite{Ghez:2008ms}.

The theoretical velocity curve can be computed as
\begin{equation}
v^{th}(r)=\sqrt{\sum_{R=IB,MB,D}v_{R}^2(r)+v_{BH}^2(r)+v_{DM}^2(r)}
\end{equation}

where 
\begin{eqnarray}
v_{BH}(r)&=&\sqrt{G\frac{M_{BH}}{r}} \nonumber \\
v_{R}(r)&=&\sqrt{\frac{G M_R}{r}}\,,R=\mbox{IB,MB,D}\nonumber\\
v_{DM}(r)&=&\sqrt{\frac{G M_{DM}}{r}}\,.
\end{eqnarray}

$M_R(r)$ is computed by integration of the mass density:
\begin{equation}
M_R(r)=8\pi a_R^3\rho_R^c(1-e^{-r/a_R})\left(1+\frac{r}{a_R}+\frac{1}{2}\left(\frac{r}{a_R}\right)^2\right)\,.
\end{equation}
Finally, the mass of the dark matter halo is computed by solving the TOV system eqs. \ref{mass} and \ref{rhoprima} for $T < T_{Deg}$ and 
eqs. \ref{mass}-\ref{rhoprima2} for $T > T_{Deg}$. Remember that the only unknowns in the solution of the TOV equations are the central density $\rho_0$ and the mass of the fermion $m_F$. 
In other words, $M_{DM}=M_{DM}(\rho_0,m_F)$. Then, the rotational velocity will depend on the choice of the dark matter central
density and the mass of the fermion.

As we have already mentioned, the rotational curve of the Milky-Way has been derived for a huge range of values, ranging from pc to kpc. The values are reported in \cite{Sofue:2013kja,2019ApJ...871..120E}.
The inner region is dominated by the luminous matter. In Fig. \ref{Fig3} we have plotted the theoretical rotational velocity curve for different scenarios:
First, by neglecting the contribution of the dark matter halo, we can see that with the values reported in Table \ref{table1}, the inner part of the rotational curve is well fitted, the inner part does not need a major contribution of dark matter. The resulting rotational velocity curve without DM is shown as a blue curve in the upper panel of Fig. \ref{Fig3}. This rotational curve arises from the effect of the matter of the BH, the inner and main bulge and the disk. Their corresponding contribution to the mass density is shown in blue lines in the lower panel of Fig. \ref{Fig3} . 
Nevertheless, the outer part needs the contribution of dark matter.  We then consider a second scenario: the full rotational velocity curve can be fitted with the inclusion of a dark matter halo that in this case will be modeled by the self gravitating structure formed by a gas of non-interacting fermions with equation of state given by eq. \ref{NR} in hydrostatic equilibrium with gravity. 
The solution for the mass function and the density depend on the central density $\rho_0$ and the mass of the fermion $m_F$, 
then, $v^{th}(r_i)=v^{th}(r_i,\rho_0,m_F)$, $r_i$ is the value of the radial coordinate where it has been reported the observed rotational velocity as reported by \cite{Sofue:2013kja,2019ApJ...871..120E}.
In order to find the value of the fermion mass and the central density of dark matter needed to fit the observed rotational curve, we have
performed a $\chi^2$ analysis given by:

\begin{equation}
\chi^2(\rho_0,m_F)=\sum_{i=1}^{67}\left(\frac{v^{th}(r_i,\rho_0,m_F)-v^{Obs}_i)}{\delta v^{obs}_i}\right)^2\,.
\end{equation}
where $v^{Obs}_i$ and $\delta v^{obs}_i$ are the observed values of the rotational velocity curve and the corresponding error as reported in \cite{Sofue:2013kja,2019ApJ...871..120E}.

\begin{figure}[t]  %
\centering
\includegraphics[scale=0.4]{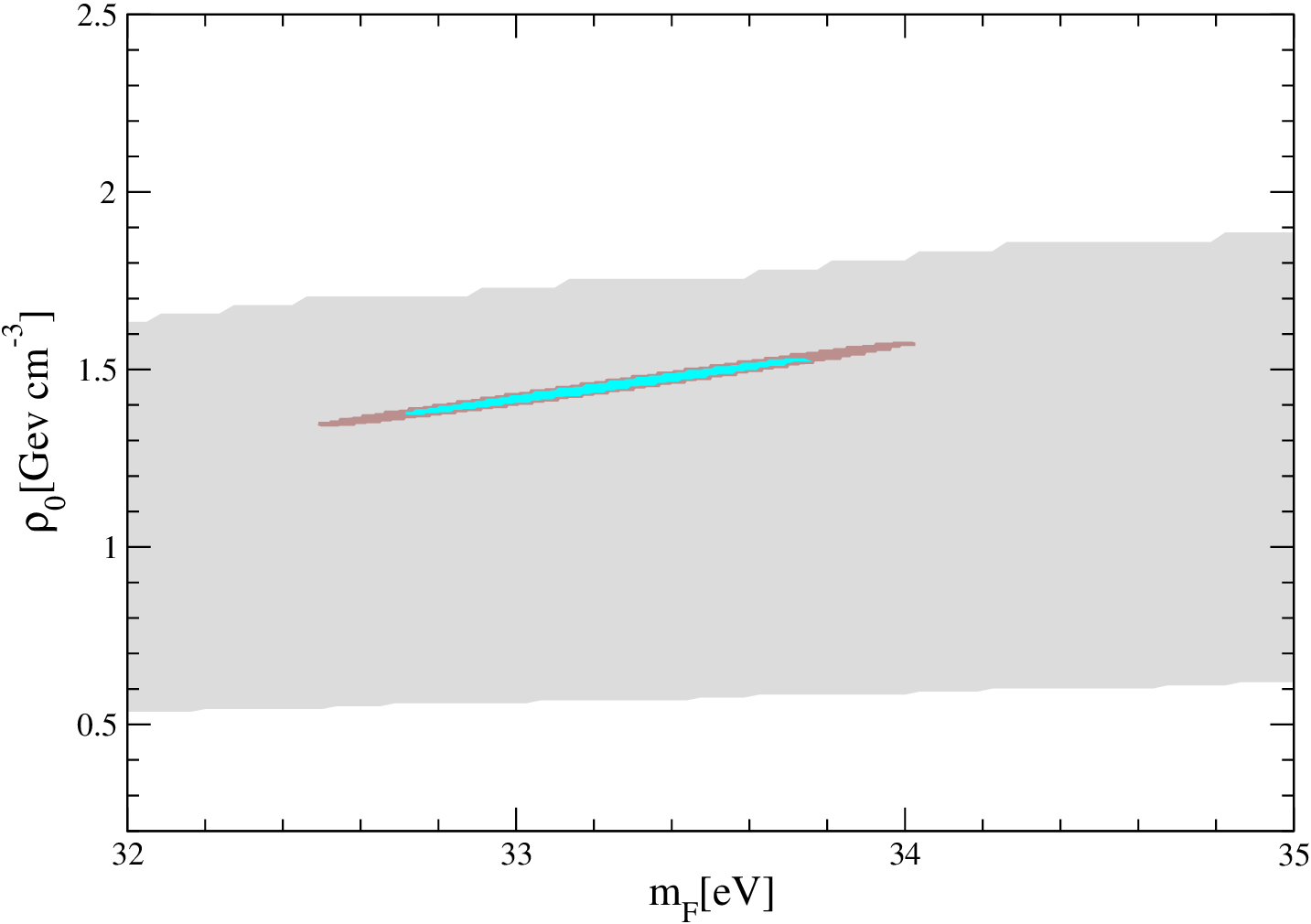}
\caption{{ Iso-curves  at $68\%$ C.L (cyan) and $95\%$ C.L. (brown) for the mass of the fermion and central densities that fit the Milky-Way rotational curve data. The allowed region in the parameter space $(m_F,\rho_0)$ that adjust  the rotational curve of the Milky-Way. In gray it is shown the region in $(m_F,\rho_0)$ that fulfills eq. \ref{salucci}.}}\label{Fig4}
\end{figure}

In Fig. \ref{Fig4} we show iso-curves  of $\Delta \chi^2=\chi^2(\rho_0,m_F)-\chi^2_{min}$ are shown . In particular, for $\Delta \chi^2=2.71$, $\Delta \chi^2=4.61$, that represent the allowed values for $m_F$ and $\rho_0$ that fits within $1\sigma$ ($68\%$ C.L.), $90\%$ C.L. and $95\%$ C.L. respectively. Thus, at $1\sigma$, the best fit point that fits the Milky-Way rotational velocity curve are: 
\begin{eqnarray}
\rho_0&=& 1.45 \pm 0.13 ~\mbox{GeV/cm}^3  \\
m_F&=& 33.23^{+0.87}_{-0.77}~\mbox{eV}\,.
\end{eqnarray}
The resulting dark matter density profile for the Milky-Way is shown as a red line in the lower panel of Fig. \ref{Fig3} and the resulting rotational velocity curve that includes the luminous matter and the DM contribution is shown in upper panel of Fig. \ref{Fig3} with a red line too. 

In order to show the differences with the standard dark matter profile and the one it is obtained with ultra light non-interacting fermions, for comparison, we have included the best fit for a Navarro-Frenk.White density profile as a black line in both figures. Observe  that the central density of the NFW dark matter halo is bigger than the corresponding central density of a semi-degenerate gas of fermions. 

The bounds obtained for $\rho_0, m_F$ with the fit to the rotational curve of the Milky-Way can be complemented with other observables. 
In particular, similar approaches for ultralight fermionic dark matter have been used to fit data from dwarf galaxies. In  \cite{Randall:2016bqw}, it is found that  a fermion with $70 < m_F < 500$ eV can fit the data of the dispersion velocities of the dwarf satellite galaxies of the Milky-Way. 
Furthermore, as we have mentioned, there is a region in the $(m_F, \rho_0)$ parameter space where the relation of a constant surface dark matter halo, i.e. eq. \ref{salucci}, is fulfilled. As it can be noted,
in order to find if the values needed to satisfy both the Milky-Way rotational curve data and the dispersion velocities of Dwarf galaxies, a combined analysis must be done and considering in both cases a semi-degenerate gas of fermions. 

We finish this section by pointing some virtues of the dark matter halo made of semi-degenerate fermions. First, the halo satisfy the scaling property expressed by eq. \ref{salucci}. This is shown in the Fig. \ref{Fig4}, where we plot the allowed region for $(m_F, \rho_0)$ that fits the rotational curve of the Milky-Way and the values of $m_F$ and $\rho_0$ that satisfies the constant density surface of the galaxies observed in \cite{Donato:2009ab,Gentile:2009bw}. As we can see, bot regions are compatible. 
{ Secondly, the local density of fermionic dark matter agrees with current observations $\rho(r=8~\rm{Kpc})=0.43\pm 0.06 ~\rm{GeV/cm}^3$ \cite{Salucci:2018hqu,deSalas:2020hbh} as we can see in Fig. \ref{fig5}.
\begin{figure}[t]  %
\centering
\includegraphics[scale=0.3]{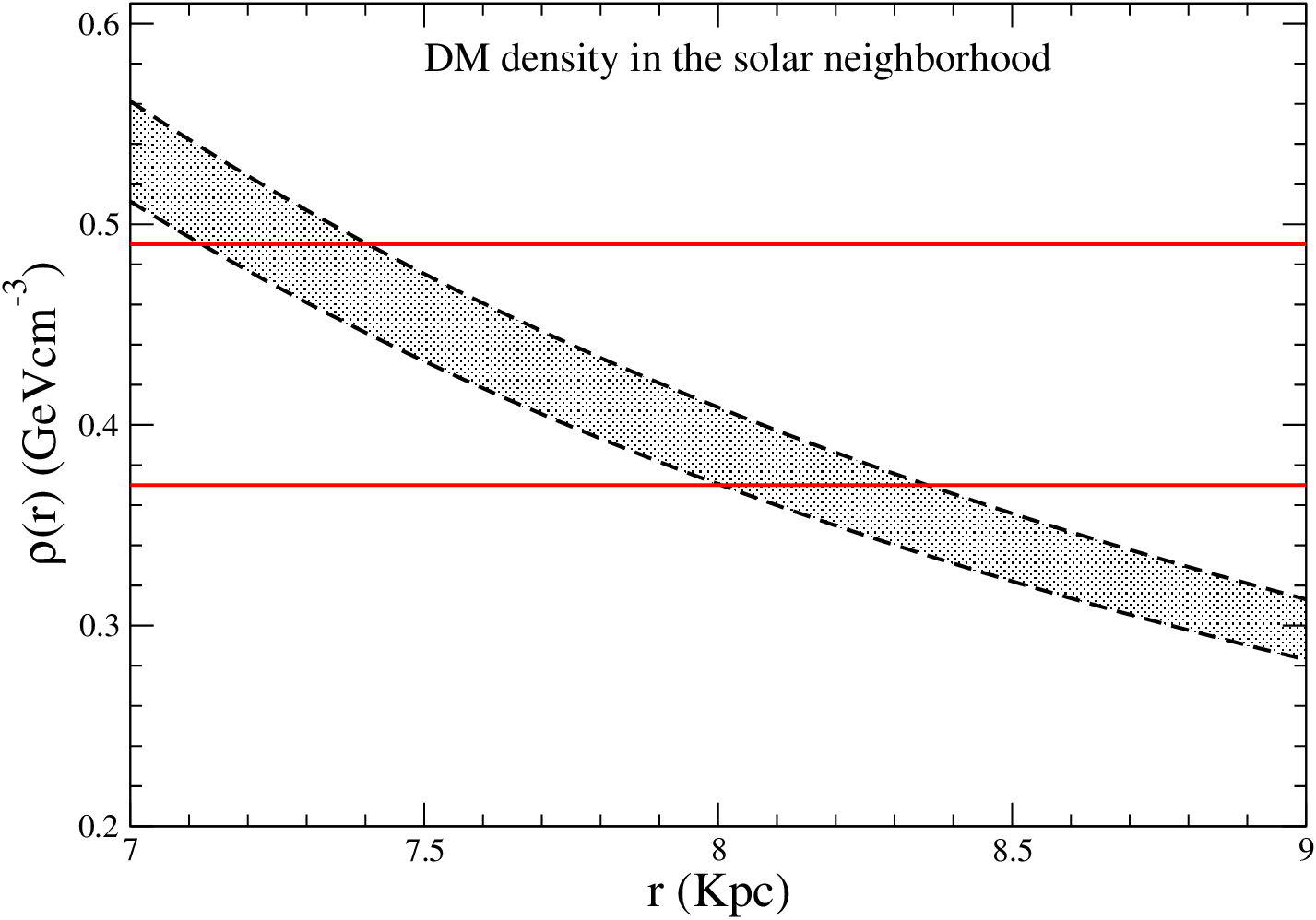}
\caption{Dark matter profile of the Milky Way halo made by fermionic dark matter with $31.5 ~\mbox{eV} < m_F < 35~$eV and $1.2 < \rho_0<1.7$ GeV/cm$^3$ in the neighborhood of the Solar System.}\label{fig5}
\end{figure}
Third: note that we have solved the TOV system for a very simple EOS that model a degenerate gas of fermions. The gas is degenerate as long as the temperature of the gas is above the degeneracy pressure. The radius where $T=T_{deg}$ implies the radius of the core. To illustrate how the energy of the fermion is distributed in the Milky Way halo, we show in {\bf Fig.  \ref{Fig6}} the Fermi energy
\begin{equation}
\varepsilon_F=\frac{1}{2m_F}\left(\frac{3\pi^2\rho(r)}{m_F}\right)^{2/3}\,,
\end{equation}
where $\rho(r)$ is the dark matter profile obtained by solving the TOV system. We can observe that the core of the galaxy consists of degenerate fermions with an almost constant Fermi energy confiming that it is the degeneracy pressure the responsible of the core of the galaxy .   
\begin{figure}[t]  %
\centering
\includegraphics[scale=0.3]{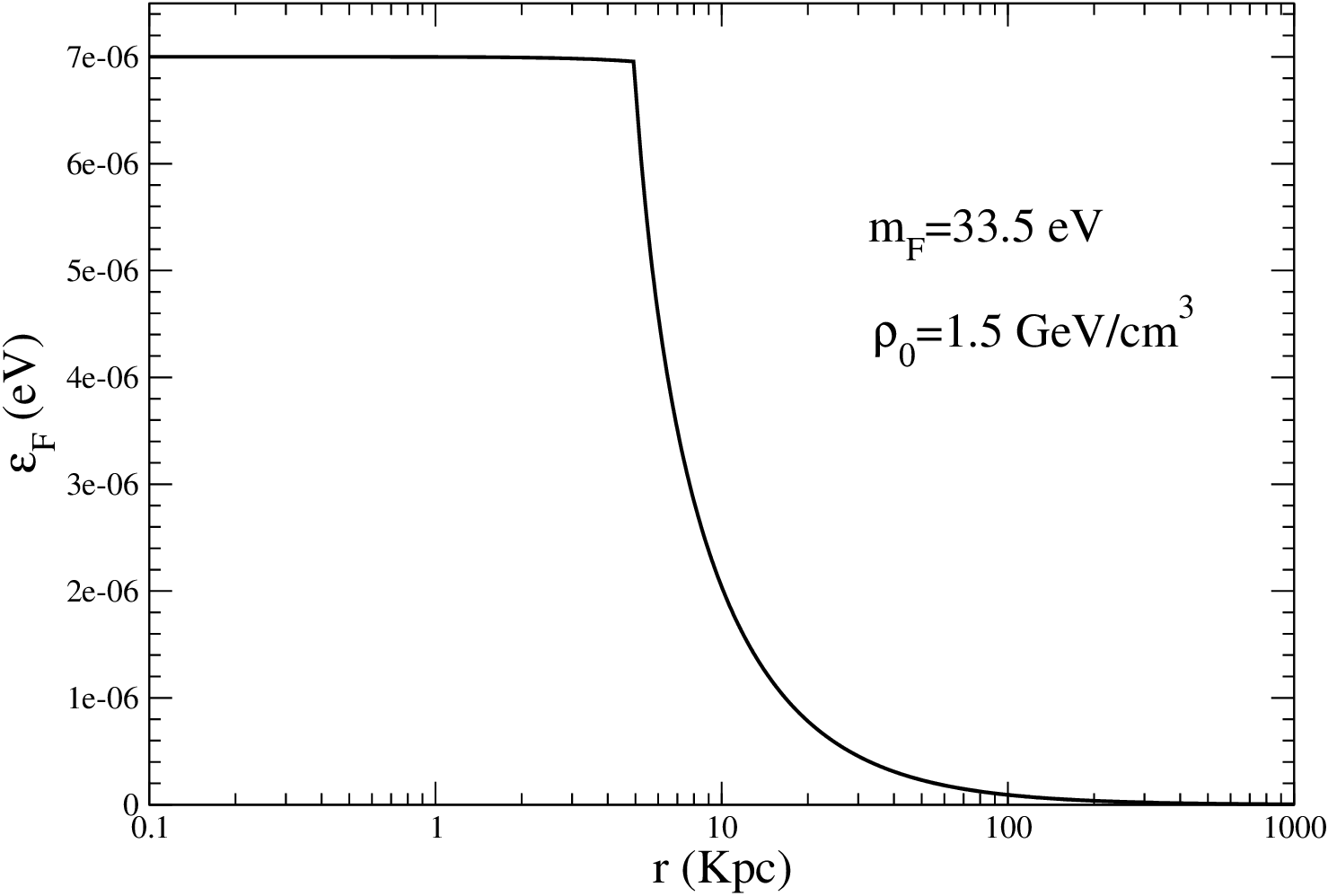}
\caption{Fermi energy $\varepsilon_F$ for the fermions for a Milky Way dark matter halo.}\label{Fig6}
\end{figure}
}

\subsection{Annihilating cross section constraints from Sgr A$^*$ data}
In this section we will use the spectral energy distribution observed from the center of the Milky-Way in order to constrain the possible annihilating cross section of this ultra-light fermion. This source of radio is known as Saggitarius A*.

\begin{figure}[t]  %
\centering
\includegraphics[scale=0.3]{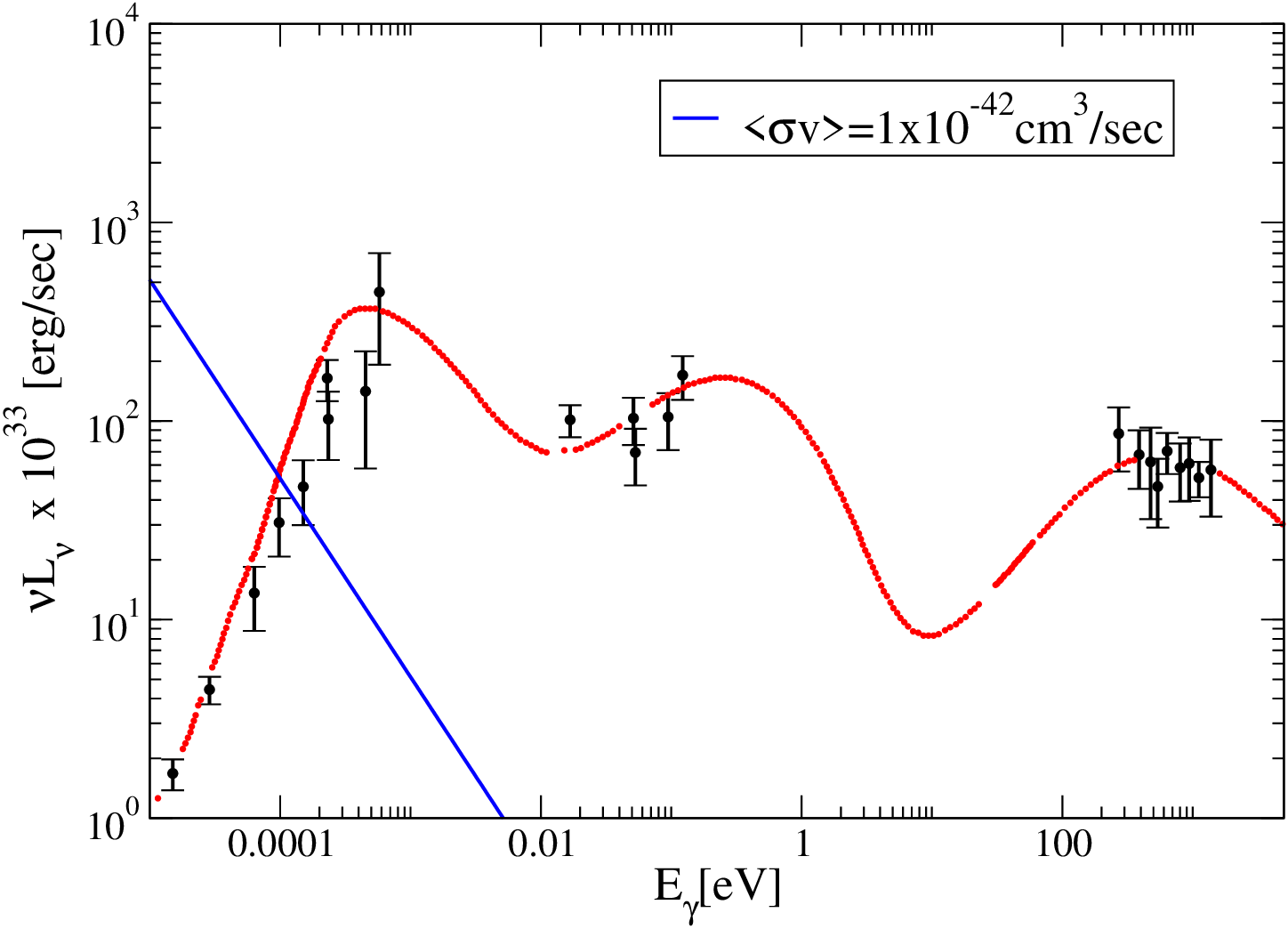}
\includegraphics[scale=0.3]{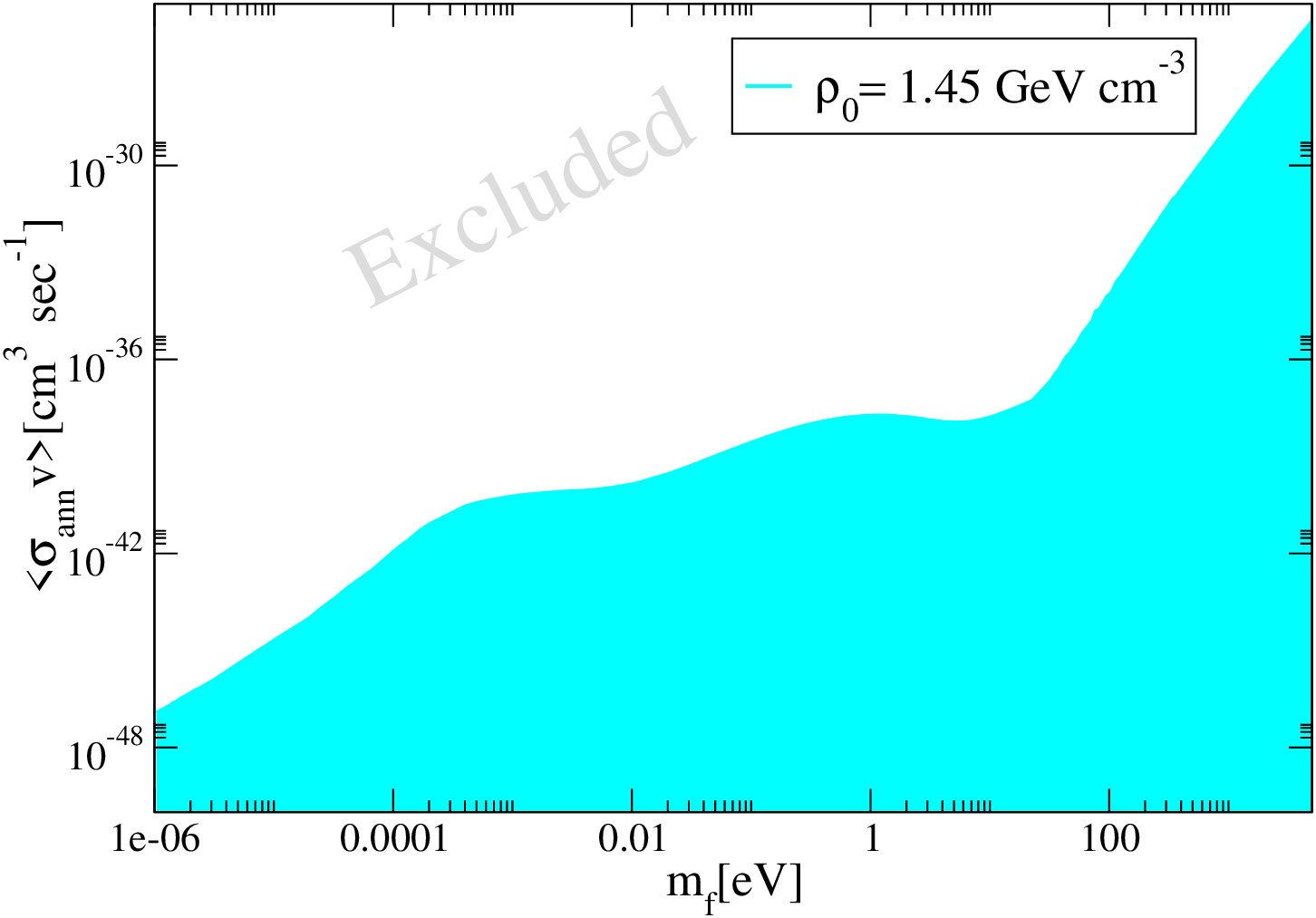}
\caption{Left: Luminosity from the galactic center. In red a model of the luminosity  \cite{Yuan:2003dc} and in blue the possible contribution 
to the spectral energy distribution of SgrA* in case that the dark matter of the Milky-Way halo is made of annihilating dark 
matter ultralight fermions. That excess in the SED will correspond to a halo with central density $\rho_0=1.45~ \mbox{Gev/cm}^3$ and a average $\langle \sigma v\rangle=10^{-42}\mbox{cm}^3/\mbox{sec}$. Right: Annihilation cross section needed to avoid an excess in the observed SgrA* spectrum energy distribution}\label{Fig7}
\end{figure}

Sagittarius A* (Sgr A*) is a compact radio source at the Galactic Center. Sgr A* is frequently monitored at all available wavelengths. First data in radio was taken in the middle 70's \cite{sgrA1974} and from that moment up to now, Sgr A* is observed in radio as reported recently in \cite{Markoff:2001ag,Zhao:2003ae}. 
There is also available data recorded in  submillimeter, near-infrared (NIR), and X-rays  \cite{Melia:2001dy,Hornstein:2002vr,Genzel:2003as}.  
In addition to observations done in the electromagnetic spectrum, there are stellar dynamical data of the stars at the galactic center that in conjuntion with the spectral energy distribution (SED) suggest that Sgr A* is a supermassive black hole.
Sgr A*'s SED can be fit with semi-analytical models \cite{Yuan:2003dc} and it is a intense source of research. For definitiveness, in Fig. \ref{Fig7} we report the data and in red a model for the SED of SgrA* as reported in \cite{Yuan:2003dc} that we will use in order to constrain possible contributions of annihilating ultra light dark matter.

We have shown that the dark matter halo of the Milky-Way can be modeled by the self-gravitating configurations made of ultralight semi-degenerate fermions.  Under the hypothesis that such fermions might annihilate themselves, there will be a extra contribution in the SED of SrgA*. Indeed, there will be a local flux of extra photons coming from the galactic center for annihilating dark matter. This flux is given by:
\begin{equation}
\phi_\gamma=\frac{1}{4\pi}\frac{N_\gamma E_\gamma}{2 m_F^2}\int \sigma v(r) \rho^2(r)dV\,. \label{fluxphotons}
\end{equation}
where $m_F$ is the particle mass, $N_\gamma$ is the number of photons per interaction, $E_\gamma$ is the photon energy, $\sigma$ is the interaction cross section, $v(r)$ is the velocity distribution of the dark matter as a function of radius and the integral is over the observed volume. We assume two photons with $E_\gamma = m_F$ per annihilation and that $v$ is independent of radius.
As an example of this possible photon  excess in the Saggitarius A* spectra energy distribution, we have computed the dark matter photon flux  that correspond to a halo configuration made of ultra light fermions 
with central density $\rho_0=47~ \mbox{Gev/cm}^3$ and an average annihilating cross section fixed as $\langle \sigma v\rangle=10^{-42}\mbox{cm}^3/\mbox{sec}$. Furthermore, we have assumed that the mass of the fermion will be equal 
to the photon energy $m_F=E_\gamma$. This contribution is plotted in a blue line the left panel of Fig. \ref{Fig7}.  

We can now use the SgrA* SED model of \cite{Yuan:2003dc} to constraint $\langle \sigma v\rangle$. In order to do so, we have to find the self-gravitating structures with $m_F=E_\gamma$ and $\rho_0=47~ \mbox{Gev/cm}^3$ by solving the TOV system for  the semi-degenerate system of fermions. Once computed the DM distribution $\rho(r)$ it is possible to compute the possible extra flux of photons with eq. \ref{fluxphotons}. The contribution to the SgrA* SED,  $\langle \sigma v\rangle$ should be smaller than the observed SgrA* spectrum and thus it is possible to constrain the annihilating cross section. 
This limit for $\langle \sigma v\rangle$ is shown in the right panel of  Fig. \ref{Fig7}.

Due to the smallness of the dark matter fermion, the number density is very high, and then, the constraints shown in Fig. \ref{Fig7} are very strong. This effect of light dark matter should be general, that is, the interaction of light dark matter with SM particles should be very small, otherwise visible effects should arise immediately due to the high number density. The lightest the dark matter is, the darkest it should become. 

\section{Conclusions}\label{section5}
We conclude that if dark matter is a ultra light, non interacting fermion, the 
dark matter halos should be modeled as the self gravitating structure supported by the quantum pressure of the fluid. In this scenario, 
halos will have a core, thus there is no "core-cusp" problem. The central densities of the halos are orders of magnitude smaller that the corresponding 
Navarro-Frenk-White profiles, thus lessening the "too big to fail" problem. 
The effective pressure of the fermions prevents DM from collapsing into small, dense structures. This quantum pressure effectively acts as a free-streaming suppression, which reduces the number of small halos that would otherwise form, resolving the discrepancy between simulations and observations of dwarf satellite galaxies around large galaxies.

In order to fit the rotational curve of the Milky-Way, the mass of the fermion is constrained to be $32.45 < m_F < 34.1$ eV and the central density of the halo lies within the range $1.33~\mbox{GeV/cm}^3 < \rho_0 < 1.58~ \mbox{GeV/cm}^3$ at $68\%$ C.L. and the resulting dark matter halo for the Milky-Way has a core radius that satisfies the  constant density surface restriction founded in \cite{Donato:2009ab,Gentile:2009bw}.
Previous studies have analyzed the dispersion velocities of dwarf galaxies and found cores of size $\sim 100$ pc for fermion dark matter with mass in the range 70 eV – 400 eV \cite{Randall:2016bqw}. It will be interesting to perform a combined analysis using the data from the rotational curve of galaxies and dwarf galaxies within the same assumptions and see if all data can be fitted with a single fermion family.
 
  The fit to the rotational curve of the Milky-Way gives a prediction for the dark matter halo density profile as shown in the red line in Figure 3. It is interesting to note that the dark matter density in the neighborhood of the Solar system, at $r=8$ Kpc is $\rho(r=8\rm{Kpc})=0.39 \pm 0.02$GeV/cm$^3$, consistent with other dark matter models.

Given the small masses of the fermions, in case that this dark matter annihilates into photons, there will be a low energy photons coming from the galactic center. 
This photons will contribute to the spectrum energy distribution and thus, by using the data of the Sgr A* SED it was possible to found constraints on $\langle \sigma v\rangle$. Constraints are very strong, 
supporting the idea that this type of fermions do not have a thermic origin. 

In summary,  data might provide strong constraints on models where dark matter is modeled by and ultra light fermion. Either by stellar dynamics data or by the low energy photons coming from
the galactic center. Combined analysis with data from Milky-Way's dwarf galaxies, structure formation can be made in order to discard this simple model of dark matter. 

\vspace{1cm}
{\bf Aknowledgements:}This work was partially support by Conacyt-SNI.

\end{document}